\documentclass[aps,amsfonts,twocolumn]{revtex4}
\usepackage{epsfig,amsmath}
\usepackage{dcolumn}                % Align table columns on decimal point
\begin{document}
\title{Weak intermolecular interactions in gas-phase NMR}
\author{\sc 
Piotr Garbacz, Konrad Piszczatowski, Karol Jackowski\footnote[1]{e-mail:kjack@chem.uw.edu.pl},
Robert Moszynski\footnote[2]{e-mail:robert.moszynski@tiger.chem.uw.edu.pl}}

\affiliation{\sl 
Faculty of Chemistry, University of Warsaw, 
Pasteura 1,02-093 Warsaw, Poland}

\author{\sc Micha{\l}~Jaszu\'nski\footnote[3]{
Author for correspondence; e-mail:michaljz@icho.edu.pl}}

\affiliation{\sl Institute of Organic Chemistry, Polish Academy of Sciences, 
Kasprzaka 44, 01-224 Warsaw, Poland}

\begin{abstract}
Gas-phase NMR spectra demonstrating the effect of weak intermolecular forces on the NMR
shielding constants of the interacting species are reported.
We analyse the interaction of the molecular hydrogen isotopomers with He, Ne, and Ar,
and the interaction in the He--CO$_2$ dimer.
The same effects are studied for all these systems in the
{\em ab initio} calculations. The comparison of the experimental and computed
shielding constants is shown to depend strongly on the treatment of the bulk susceptibility
effects, which determine in practice the pressure dependence
of the experimental values. Best agreement of the results is obtained
when the bulk susceptibility correction in rare gas solvents is evaluated from
the analysis of the He-rare gas interactions, and when
the shielding of deuterium in D$_2$--rare gas systems is considered.
\end{abstract}

\date{\today}
\maketitle

\newpage

\section{Introduction}
\label{sec1}
The importance of intermolecular interactions in physics, chemistry,
and biology does not need to be stressed.
Intermolecular potentials determine the properties of non-ideal
gases, (pure) liquids, solutions, molecular solids, and the behavior of
complex molecular ensembles encountered in biological systems. They
describe the so-called non-bonded contributions, as well as the special
hydrogen bonding terms, that are part of the force fields used in
simulations of processes such as enzyme-substrate binding, drug-receptor
interactions, etc. A few examples showing important applications
of intermolecular potentials include the
Monte Carlo and molecular dynamics simulations of biological systems,
studies of processes in the earth's atmosphere, or
interstellar chemistry.

Also the NMR spectra, in particular the observed chemical shifts, depend
not only on the molecular structure but also on the intermolecular
forces. The changes due to the environment are difficult to interpret
theoretically and make the comparison of the computed and observed
spectra unreliable. The role of the intermolecular forces is
undoubtedly the largest in the condensed phase, and much smaller in dilute
gas-phase solutions. Moreover, it is particularly small if we analyse
a system where only weak van der Waals intermolecular forces play
a significant role. In this work, we describe gas-phase NMR spectra
for such systems, analyse the dependence of the observed shielding
constants on the intermolecular forces, and present {\em ab initio} 
calculations which describe this dependence.

Early NMR studies in the gas phase were reviewed by Rummens~\cite{fhar-rev},
another review has been written in 1991 by Jameson~\cite{cjjcr91}.
However, the role of the intermolecular interactions in the gas phase was 
almost exclusively interpreted on the basis of binary collision gas model 
introduced by Raynes, Buckingham,
  and Bernstein~\cite{wtradbhjbjcp36}.
In this RBB model the change in the shielding constant
is qualitatively described as a sum of contributions due to
the bulk susceptibility, neighbor-molecule magnetic anisotropy,
polar effects, and van der Waals effects.
At present, by applying state-of-the-art methods of quantum chemistry
we should be able to
predict accurately the small changes of the shielding constants due
to weak intermolecular forces.  
For the first time this should be possible within
an {\em ab initio} approach, which is in principle more
reliable than the standard
methods used to describe for instance the solvent effects in liquids,
such as various polarizable continuum models based on classical
approximations.
Therefore, a study of  gas-phase model systems has a specific advantage
for the comparison between experiment and theory. 

Theoretical studies of the interaction-induced changes in the NMR
parameters are scarce, and mostly restricted to supermolecule calculations of the
interaction-induced shielding constants and spin-spin coupling
constants; see, e.g. Refs.
\cite{abmjthkrcpl250,mpjscp234,mpjscp248,kjmwmpjsjpca104,mparmp100,mbmkolmvgmdrsjcc20,mhplnrjjjvjcp121,mhplmihjajjvjcp127}
for typical applications. To our knowledge only one paper~\cite{abmjthkrcpl250} 
analysed (comparing the theory with the numerical results) the asymptotic long-range behavior of the 
shielding constant and its anisotropy in a dimer.
Most of the papers reporting {\em ab initio} calculations
of the NMR parameters that could directly be compared with the gas phase
NMR experiment were devoted to studies of atom-atom 
interactions~\cite{aamjarjcp126,mparmp100,mbmkolmvgmdrsjcc20,mhplnrjjjvjcp121,mhplmihjajjvjcp127}  
(this is in sharp contrast with the electric properties
of molecular complexes 
for which a general long-range theory and applications to the optical and
dielectric properties of gases are available 
\cite{tgahrmpeswavdamp89,rmtgahavdacpl247,arscdmjlcbfchmp104}).
There are very few
{\em ab initio} studies of the NMR effects of weak interactions between a molecule and an atom
or two molecules. The whole property surface has been
computed for the interactions in the 
C$_{2}$H$_{2}$--He and C$_{2}$H$_{2}$--H$^+$ complexes~\cite{mpjscp248}, but no
Boltzmann averaging has
been performed;
NMR properties were also examined~\cite{mpjscp234}
for the optimized geometries of other binary complexes of acetylene.

Also on the experimental side not too much has been done. 
The effects of weak
intermolecular interactions on NMR shielding 
of $^1$H, $^2$H and $^3$He in gases are small
and buried in the much larger bulk susceptibility
correction, therefore a detailed analysis of such systems is
practically impossible and the RBB model has mostly been used. 
We recall here that for $^3$He the effect of the weak interactions is 
particularly small and difficult to observe 
(see for instance the study of gas-to-liquid shifts~\cite{rspdrkmjjjmra101}).
For $^{21}$Ne the
precision of the NMR measurements is limited because the
magnetically active isotope has a large nuclear quadrupole moment;
for argon the only magnetically active $^{39}$Ar 
isotope is radioactive.
On the other hand, for $^{129}$Xe the effects are very large.
They have been observed in the xenon dimer and nearly quantitative agreement of
theory with experiment was reached in
state-of-the-art {\em ab initio} calculations~\cite{mhplnrjjjvjcp121,mhplmihjajjvjcp127}.
Also density functional theory (DFT) calculations for Xe-rare gas dimers yield satisfactory 
agreement with experimental data, 
see Refs.~\cite{cjjakjsmcjcp62,cjjdnsacdjcp118}. Most recently, the chemical shift
of Xe dissolved in liquid benzene was studied in 
the calculations combining the DFT methods with 
the classical molecular dynamics~\cite{sspkrmjhpbmstca129}.  
However, there are no similar studies of atom-molecule systems applying 
well established state-of-the art wavefunction methods and comparing the results 
with known experimental data.

In this paper we fill this gap and report a joint experimental and 
theoretical study of the gas phase shielding constants in the mixtures
of atomic and molecular gases. 
We study the effects resulting from
the weak interactions between a molecule and an atom
in series of model systems: H$_2$--He, H$_2$--Ne and H$_2$--Ar dimers
and their deuterium-substituted isotopomers and in
He--CO$_2$. For the
selected magnetically active nuclei---$^1$H, $^2$H, $^3$He and $^{13}$C---
we observe the dependence of the NMR spectrum on the density
of the solvent gas, which enables next a comparison of the {\em ab initio} and experimental 
results.
In the
analysis of the NMR spectra we take into account the bulk 
susceptibility correction, 
dependent on 
the magnetizability of the medium and on the shape of the NMR 
sample~\cite{rspdrkmjjjmra101,japwgshjbbook}. 
In the case of the weakly interacting systems which we study  
this correction dominates in the density dependence of the spectrum
and its proper description is essential
when we extract the information
on the role of the intermolecular interactions from the experimental data 
and compare the experimental and computed quantities.

The plan of this paper is as follows. We start with  the virial
expansion of the shielding constant in terms of the gas density
and discuss all quantities needed on the route from the theory to a
direct comparison with the experiment. This is thoroughly discussed
in sec. \ref{sec2}. The details of the computational
procedures adopted in the {\em ab initio} calculations, fitting
of the interaction potential and shielding surfaces, and
some numerical integration procedures will be discussed in sec. \ref{sec3}.
The experiment is described in detail in sec. \ref{sec4}. The results of
the measurements and calculations are reported and compared in sec. \ref{sec5}.
Finally, sec. \ref{sec6} concludes our paper.

\section{Shielding constants in the gas-phase solutions}
\label{sec2}
For a binary mixture of a gas $A$, containing the nucleus $X$ whose shielding $\sigma^A(X)$ is observed, and gas $B$ as the solvent, $\sigma^A(X)$ can be expressed as~\cite{wtradbhjbjcp36}:
\begin{equation}
        \sigma^A(X) = \sigma_0^A(X) + \sigma_{1}^{AA}(X) \rho_A + \sigma_{1}^{AB}(X) \rho_B
\label{eq:AB}
\end{equation}
where $\rho_A$ and $\rho_B$ are
the densities of $A$ and $B$, respectively, and $\sigma_0^A(X)$ is 
the shielding in the zero-density limit. All higher terms in Eq.~(\ref{eq:AB}),
which represents a truncated virial expansion,
can safely be neglected if 
the experimental dependence of the shielding on the density is linear. The 
coefficients $\sigma_{1}^{AA}(X)$ and $\sigma_{1}^{AB}(X)$ are then
the only terms responsible for the medium effects. 
They contain the bulk susceptibility corrections, $\sigma_{{\rm 1bulk}}^A$ and $\sigma_{{\rm 1bulk}}^B$, and 
the terms directly taking account of the intermolecular interactions during 
the binary collisions of the $A-A$ and $A-B$ molecules: $\sigma_1^{A-A}(X)$ and $\sigma_1^{A-B}(X)$, respectively. The shielding parameters in 
Eq.~(\ref{eq:AB}) are temperature dependent and for this reason all 
the present measurements are performed at 
the constant temperature of 300 K. Moreover, in the experiments 
the density of $A$, $\rho_A$, is always kept very low in order to eliminate 
the solute-solute molecular interactions and Eq.~(\ref{eq:AB}) can be simplified to:
\begin{equation}
        \sigma^A(X) = \sigma_0^A(X) + \sigma_{1}^{AB}(X) \rho_B
\label{eq:B}
\end{equation}
where
\begin{equation}
        \sigma_{1}^{AB}(X)  = \sigma_{{\rm 1bulk}}^B + \sigma_1^{A-B}(X) .
\label{eq:ABA-B}
\end{equation}

\begin{figure}
\begin{center}
\includegraphics[scale=0.3]{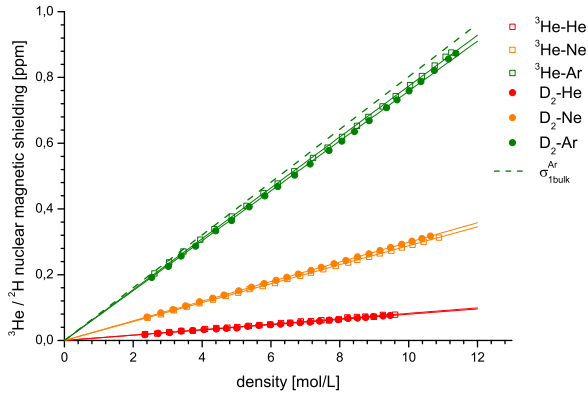}
\end{center}
\caption{The observed density-dependent $^3$He shielding of 
atomic helium and $^2$H shielding of deuterium 
in gaseous solutions (for comparison, the best estimate of the  $\sigma_{{\rm 1bulk}}^{{\rm Ar}}\, \rho_{{\rm Ar}}$ contribution 
is shown).}
\label{figd2}
\end{figure}

Fig.~\ref{figd2} displays, as an example, the dependence of the helium and 
deuteron magnetic shielding (given with respect to the isolated systems) on 
the density of the rare gas solvent in gaseous solutions. 
The plots in Fig.~\ref{figd2} are linear,  which proves that 
Eq. (\ref{eq:B}) is a valid approximation and allows 
the determination of the $\sigma_0^A(X)$ and $\sigma_1^{AB}(X)$ shielding parameters.
The part of the shielding constant $\sigma^A(X)$ which is exclusively
due to pair intermolecular interactions between the solute and solvent molecules
is given by $\sigma_1^{A-B}(X)$. An inspection of Eqs. (\ref{eq:B}) and (\ref{eq:ABA-B}) 
shows that $\sigma_1^{A-B}(X)$ can be extracted from the experimental results once
the measured shielding constant becomes linear in the gas density $\rho_B$, and if the bulk
susceptibility correction, $\sigma_{{\rm 1bulk}}^B$, is known.

In the experiment it is not easy to measure the gas number
density $\rho$, but rather the pressure $p$. Therefore, the following
form of Eq. (\ref{eq:B}) was used:
\begin{equation}
\sigma^{A}(X)=\sigma_0^{A}(X)+\sigma_{1p}^{AB}(X)p.
\label{eq:Bp}
\end{equation}
For an ideal gas Eqs. (\ref{eq:B}) and (\ref{eq:Bp}) are equivalent,
and the coefficients $\sigma_{1p}^{AB}(X)$ and $\sigma_{1}^{AB}(X)$ are
inter-related by the following simple expression:
\begin{equation}
\sigma_{1p}^{AB}(X)=\sigma_{1}^{AB}(X)/k_BT.
\label{rel}
\end{equation}
In general Eq. (\ref{rel}) is not valid since the pressure depends on the
gas number density in a more complicated way:
\begin{equation}
p = k_BT\rho + B_2(T)\rho^2 + B_3(T)\rho^3 + \cdots,
\label{pvir}
\end{equation}
where $B_2(T)$ and $B_3(T)$ are the second and third thermodynamic virial coefficients, respectively.
Assuming that the NMR active molecules
are infinitely diluted in the bath, $B_2(T)$ exclusively depends on the pair interactions
between the molecules in the bath.
The third virial coefficient
$B_3(T)$ also depends on the non-additive three-body interactions in the bath.
We assume that we are dealing with an infinitely diluted solutions.
In such a case we can assume that the concentration
of the solute is very small and that the contribution of the partial pressure of
the solute to the
total pressure is negligible. This means that the thermodynamics of the system is described
by Eq. (\ref{pvir}) with the characteristic coefficients of the solvent used in
the experiment, while
Eq. (\ref{eq:B}) describes the change of the shielding constant due to binary
collisions of the NMR active molecule with the bath molecules.
It is worth noting at this point that the virial expansions (\ref{pvir})
and (\ref{eq:B}) follow from
the theory, and that in the virial expansions the gas number density appears as the
variable of the power series.

A precise evaluation of
$\sigma_{{\rm 1bulk}}^B$, the bulk susceptibility correction, is particularly important in
the present work,
because for all the nuclei the total change of the shielding 
due to
intermolecular interactions
in the gas phase is very small. We consider first
the determination of
$\sigma_{{\rm 1bulk}}^B$ terms from the available $\chi_{\rm M}$ ---molar 
magnetic susceptibilities of gases--- and applying
the standard formula for a infinitely long cylindrical tube parallel to
the external magnetic field ~\cite{rspdrkmjjjmra101,japwgshjbbook,beckerbook}:
\begin{equation}
\sigma_{{\rm 1bulk}} =  - \frac{4 \pi}{3} \;  \chi_{\rm M} \;
\label{eq:4pi}
\end{equation}
where $\chi_{\rm M}$ is given in ppm cgs and $\sigma_{{\rm 1bulk}}$ in
ppm mL/mol. 
The macroscopic molar magnetic susceptibility $\chi_{\rm M}$ is
for closed shell systems proportional to the microscopic molecular magnetizability
(1~ppm cgs corresponding to 
16.60529 $\times 10^{-30}$\ J\,T$^{-2}$).
Equation~(\ref{eq:4pi}) may be applied for the cylindrical geometry of the sample,
and assuming that the molecules of the solvent do not interact. 
However, this assumption is not always true, the cylinder is
not infinite, and in such an approach various additional corrections are 
undoubtedly necessary
to get realistic values of $\sigma_{{\rm 1bulk}}^B$ (see for instance
Seydoux {\it et al.}~\cite{rspdrkmjjjmra101}). 
We use a different approach to determine $\sigma_{{\rm 1bulk}}^B$,
which
will be discussed in detail in section~\ref{sub5}.

Theoretical determination of $\sigma_1^{A-B}(X)$ requires two steps:
{\em ab initio} calculations of the interaction potential and 
interaction-induced shielding constant for the binary complex \mbox{$A-B$},
and the average of the latter quantity with the Boltzmann factor
depending on the interaction potential. The interaction potential
$V$ is given by the standard expression:
\begin{equation}
V=E_{AB}-E_A-E_B,
\label{vint}
\end{equation}
where $E_{AB}$, $E_A$, and $E_B$ are the energies of the collisional
dimer $A-B$, and of the solvent ($A$) and solute ($B$) molecules,
respectively. The interaction-induced shielding constant
$\sigma_{\rm int}^{A-B}(X)$ is given by:
\begin{equation}
\sigma_{\rm int}^{A-B}(X)=\sigma_0^{A-B}(X)-\sigma_0^A(X),
\label{sint}
\end{equation}
where $\sigma_0^{A-B}(X)$ and $\sigma_0^A(X)$ are the shielding constants
of the nucleus $X$ in the dimer $A-B$ and in the solute molecule $A$, respectively.
Finally, $\sigma_{1}^{A-B}(X)$ appearing in Eq. (\ref{eq:ABA-B}) is defined
as:
\begin{equation}
\begin{split}
&\sigma_{1}^{A-B}(X)=\\&\int\int\int\sigma_{\rm int}^{A-B}(X)
\exp\left(-\beta V(\omega_A,\omega_B,R)\right)R^2{\rm d}R{\rm d}\omega_A
{\rm d}\omega_B,\
\end{split}
\label{boltz}
\end{equation}
where $\omega_A$ and $\omega_B$ denote the two sets of the angles
specifying the orientations of the monomers $A$ and $B$,  $R$ is
the distance between the centers of mass of the monomers,
$\beta=(k_BT)^{-1}$,
$k_B$ is the Boltzmann constant, and $T$ is
the temperature in Kelvin. We note that in general the calculation of 
$\sigma_{1}^{A-B}(X)$ is not an easy task. For rigid molecules $A$
and $B$ it requires a six-dimensional integration over five angles
and one distance. For systems considered in the present paper the
integral of Eq. (\ref{boltz}) reduces to a two-dimensional integral
that can easily be evaluated. The details of the computational
procedures adopted in {\em ab initio} calculations, fitting, and
numerical integration will be discussed in the next section.

\section{Computational approach}
\label{sec3}
\subsection{Ab initio calculations}
\label{sub1}
In all calculations the bond lengths of the interacting subsystems
were kept fixed at their experimental 
geometries.
We report below the results for H$_2$ obtained with the H--H distance
fixed at 
$r$(HH) = 1.449~a$_0$~\cite{wklwjcp41}. In test calculations
we have verified that practically the same results are obtained
using noticeably smaller values of $r$(HH).
Thus, we can compare the same set of {\em ab initio} results with the experimental
data for different isotopomers of the hydrogen molecule.
For He--CO$_2$, following the previous studies of the
potential energy
surface~\cite{tkrmftjmlbbhjbpeswjcp115}, we have used
$r$(CO)= 2.1944~a$_0$, 
an experimental value deduced from the microwave spectra.

All calculations of the energies and of the shielding constants
have been performed with the coupled cluster method
restricted to single, double, and noniterative triple
excitations, CCSD(T).
The NMR shielding constants and
the magnetizabilities were obtained by
applying the coupled cluster linear response theory~\cite{jgjfsjcp102,jgjfsjcp104}.
Gauge-including atomic orbitals,
GIAO's~\cite{fljpr8,kwjfhppjacs112}, were used in all
calculations of the magnetic properties, and
we have systematically corrected all the interaction-induced
changes in the energies and in the shielding constants by eliminating
the basis set superposition error, i.e. all calculations for the
monomers were done in the full basis of the dimer.
We have used the d-aug-cc-pVXZ basis sets~\cite{rakthdrjhjcp96}; 
d-aug-cc-pVQZ for the smallest H$_2$--He system, and
the d-aug-cc-pVTZ  basis set for the larger
H$_2$--Ne, H$_2$--Ar, and He--CO$_2$ dimers.
The calculations of the energies and shielding constants
were performed using the ACES~II~\cite{aces2:2006}
program, while the magnetizabilities were computed using
the more recent CFOUR program~\cite{cfour:09}.

\subsection{Interaction potentials}
\label{sub2}
For two systems we used the available
fitted interaction potential energy surfaces: for H$_2$--Ar taken from
Ref.~\cite{hlwksbjrmsrjcp98} and for He--CO$_2$ taken from
Ref.~\cite{tkrmftjmlbbhjbpeswjcp115}. These potentials were obtained
from the symmetry-adapted perturbation theory (SAPT) calculations (see
Refs.~\cite{bjrmkscr94,rm:rev}
for a review of the SAPT methodology and of the accuracy of the SAPT potentials).
These potentials
were shown to reproduce the high-resolution infrared spectra of
the H$_2$--Ar~\cite{rmbjpeswavdacpl221,fmrmjcp109} and
the He--CO$_2$~\cite{tkrmftjmlbbhjbpeswjcp115}
van der Waals complexes. More importantly, they also reproduce very
accurately the thermodynamic (pressure) virial coefficients
\cite{rmtktgahpeswavdabspjch72}.
For other systems the interaction potential $V(R,\theta)$ was
calculated by the supermolecular method according to Eq. (\ref{vint}).

We use spherical coordinates defined with respect to the center of mass of
the molecule.
Calculations were performed for several angles $\theta$ ranging from 0 to 180$^\circ$ and for several radial distances  $R$.
For each angle $\theta$ radial dependence of the
interaction potential $V$ was fitted with the function:
\begin{equation}
\begin{split}
V_\theta(R)&=e^{-\alpha(\theta) R}(A_0(\theta)+A_1(\theta) R+A_2(\theta) R^2)\\
&-\frac{C_6(\theta)}{R^6}-\frac{C_8(\theta)}{R^8}\,,
\end{split}
\label{vfit}
\end{equation}
where $\alpha$, $A_0$, $A_1$, $A_2$, $C_6$, and $C_8$ were adjusted to fit the
computed points at a given angle $\theta$.
We note parenthetically that odd powers of $R^{-1}$ do not appear in the long-range
asymptotics of Eq. (\ref{vfit}) because the H$_2$ and CO$_2$ molecules
are centrosymmetric.
Next, interpolation was used to obtain the full interaction energy surface.
The points calculated for a given radial distance $R$ from each $V_\theta$
fit were interpolated with a third-order polynomial in $\theta$. This procedure leads to a fitted/interpolated
interaction energy surface $V(R,\theta)$, which was used in further calculations.

\subsection{Shielding constants}
\label{sub3}
The same technique was applied to obtain the $\sigma_{\rm int}^{A-B}(R,\theta)$ surface.
For each angle the radial dependence of $\sigma$ was fitted to the following function:
\begin{equation}
S_\theta(R)=e^{-\alpha(\theta) R}\sum_{k=0}^NA_k(\theta) R^k-\sum_{m\in M}\frac{C_m(\theta)}{R^m},
\end{equation}
where all the parameters appearing on the r.h.s. of the expression above were
adjusted to fit the computed values.

For the hydrogen atom in H$_2$--He, H$_2$--Ne, and H$_2$--Ar a modification to 
the procedure described above was introduced.
Since the interaction energy surface for these systems is symmetric we were allowed to use symmetrized
$\sigma$-surface $\bar\sigma(R,\theta)$ calculated as an arithmetical average of 
interaction-induced shifts for both H nuclei.
This improved the accuracy of the further integration of the $\sigma_{\rm int}^{A-B}$ function with the Boltzmann factor.

\subsection{Final integration}
\label{sub4}
To obtain the final result one has to calculate for the temperature of interest 
the Boltzmann average of 
$\sigma_{\rm int}^{A-B}(R,\theta)$:
\begin{equation}
\label{eq:Boltz}
\begin{split}
&\sigma_1^{A-B}  =\\
&\int_0^\infty \!\!\!{\rm d}R \int_0^\pi\!\!\! {\rm d}\theta
\int_0^{2\pi}\!\!\!{\rm d}\phi R^2\sin\theta \; \exp\left(-\beta V(R,\theta)\right) \sigma_{\rm int}^{A-B}(R,\theta).
\end{split}
\end{equation}
Due to the axial symmetry of the considered systems the integration over $\phi$
gives 2$\pi$. Integrations over $R$ nad $\theta$ were done numerically with
the {\sc Mathematica}~\cite{Mathematica7}  package.
First, for each angle the radial integration was performed. The integration range [0,$\infty$[ was substituted by [$R_{\rm min}$, $R_{\max}$]
with properly defined $R_{\rm min}$ and $R_{\rm max}$. The $R_{\rm min}$
value was chosen
to ensure that $V(R_{\rm min},\theta)$ was positive and large enough to make the Boltzmann factor close to zero.
The value of $R_{\rm max}$ was chosen in such a way that
$\sigma(R_{\rm max},\theta)$ was almost zero at $R_{\rm max}$, independent of the
angle $\theta$.
This choice leads to $R_{\rm min}$ = 2~a$_0$ and
$R_{\rm max}$ = 300~a$_0$  for H$_2$--He, for other systems the
required integration range is smaller and within the same limits.
The results obtained from the radial integration for each angle were interpolated with third-order function and this function was integrated over $\theta$.

\subsection{Bulk susceptibilities}
\label{sub5}
To estimate the bulk susceptibility correction (BSC) we first used new values of
magnetizabilities obtained from CCSD(T) calculations.
Using the d-aug-cc-pVQZ basis set we obtain for CO$_2$
at the experimental geometry
--22.254~ppm~cgs, with the basis set error estimated to be
smaller than 0.2~ppm~cgs. This value is consistent with the
results of Ref.~\cite{krprtmjjpca104}
and confirmed by new CCSD(T)/d-aug-cc-pCVQZ-unc calculations for the Ne and Ar atoms, which
give --7.601 and --20.610~ppm~cgs, also in agreement with Ref.~\cite{krprtmjjpca104}.
The values of $\chi_{\rm M}$ and the corresponding bulk susceptibility effects derived
from  Eq.~(\ref{eq:4pi})  are given in Table~\ref{tab:magn}.
These values may only be considered as a crude approximation to
the real $\sigma_{{\rm 1bulk}}^B$ quantities. First, because the geometric factor is unable to reproduce accurately
the susceptibility corrections in nuclear shielding; this problem was frequently discussed in 
the literature from
the early days of NMR~\cite{aabregjcp26,aabjmsp5}.
Secondly, we have used
a special high-pressure tube, cf. sec.~\ref{sec4},
which was not spinning and this may induce nonnegligible 
unknown effects.

\begin{table}[ht]
\caption{{Solvent gas $(B)$ magnetizabilities (ppm cgs) and  bulk susceptibility corrections (ppm mL/mol)}}
\label{tab:magn}
\begin{center}
\begin{tabular}{l d d d d r}
\hline
\hline
 & \mbox{He}  &  \mbox{Ne} &  \mbox{Ar} &  \mbox{CO$_2$} \\
\hline
$\chi_{\rm M}$ \footnotemark[1] & -1.8915 & -7.601 & -20.610  & -22.254 \\
$\sigma_{{\rm 1bulk}}^B$\footnotemark[2]  &7.923  & 31.839 &86.333 &  93.217\\
$\sigma^{HeB}_{\rm 1}\, $\footnotemark[3]  &8.29(12)  & 28.79(6) & 77.36(30) &         &        \\
$\sigma_{{\rm 1bulk}}^B$\footnotemark[4]  &8.62(12)  & 29.56(6) & 80.26(30) &         &        \\
\hline\hline
\end{tabular}
\end{center}

\footnotetext[1]{For consistency with Eq.~(\ref{eq:4pi}) we use ppm cgs units.}
\footnotetext[2]{Calculated from Eq.~(\ref{eq:4pi}).}
\footnotetext[3]{Total effect observed in $^3$He--B interaction.}
\footnotetext[4]{Determined from the total $^3$He--B interaction and the computed interaction-induced
$\sigma_1^{He-B}$ coefficients.}
\end{table}

Since a precise determination of the BSC value according
to Eq.~(\ref{eq:4pi})
is impossible,
we have applied our own experimental approach
to estimate the bulk susceptibility corrections.
It is well known that molecular interactions between the atoms of rare gases disturb
the $^3$He shielding only to very small extent~\cite{rspdrkmjjjmra101}. Moreover,
a description of such interactions is available from the
theoretical studies of the shielding in these gas mixtures~\cite{aamjarjcp126}. In
the present work we have measured the density dependent $^3$He shielding in helium, neon, and argon gases. It gave us
the $\sigma_{1}^{HeHe}$, $\sigma_{1}^{HeNe}$ and $\sigma_{1}^{HeAr}$  coefficients of
Eq.~(\ref{eq:B}), which were used next in Eq.~(\ref{eq:ABA-B})
together with
the interatomic interaction coefficients, to obtain 
$   \sigma_{{\rm 1bulk}}^B$ as $\sigma_{1}^{AB}(X) -  \sigma_1^{A-B}(X)$.
We have used the values of 
$\sigma_1^{He-He}$, $\sigma_1^{He-Ne}$ and $\sigma_1^{He-Ar}$,
based on the theoretical results
 of Ref.~\cite{aamjarjcp126}:
--0.328, --0.776 and --2.901~ppm mL/mol, respectively
(another available value of 
$\sigma_1^{He-He}$, derived from the full configuration interaction calculations, but with a smaller basis set,
 is equal to --0.353~ppm mL/mol~\cite{mparmp100}).
In this way we determined
the final values of the bulk susceptibility effects in
the present NMR experiments, shown in Table~\ref{tab:magn}.
Finally, we note that the problems related to precise
determination of the bulk susceptibility effects are known, they
have been recently analysed~\cite{mdpbjctc6,rehjmr178} and
discrepancies of the order of $\approx$10\% between the computed and
experimental data have been observed~\cite{mdpbjctc6}.

\section{Experiment}
\label{sec4}
The $^1$H, $^2$H, $^3$He and $^{13}$C NMR chemical shifts were measured on a Varian INOVA 500 
spectrometer at 300~K operated at 500.61, 76.85, 381.36 and 125.88~MHz, respectively. 
$^2$H and $^{13}$C spectra were acquired with a standard two channel Varian switchable 5 mm probe,
 while $^3$He and $^1$H spectra in
the self reconstructed helium
probe~\cite{kjmjbkmwjmr193}.
Nitromethane-d$_3$ was used for a lock system when the $^1$H, $^3$He, and $^{13}$C NMR measurements were carried out. The $^2$H experiments required a high-band lock operating on
the proton signal of liquid tetramethylsilane (TMS). For this purpose a special set of coaxial
glass capillaries was prepared and the
same set was also used for the external referencing of all the chemical shifts. The set of capillaries contained nitromethane-d$_3$ in
the outer chamber and pure liquid TMS in the inner container.
The capillaries were placed in a special non-spinning NMR tube which
was used for all our measurements.
The tube was made of zirconia and equipped with a metal valve for gas
filling at high pressure (Daedalus Innovations, USA).

The described sample setup was complex, the zirconia tube with the 
capillaries affects the external magnetic field, and  
therefore we could not apply Eq.~(\ref{eq:4pi}) in our experimental work
(see also the discussion in Ref.~\cite{rspdrkmjjjmra101}). 
We have bypassed the problem of  bulk susceptibility corrections 
performing analogous measurements of 
$^3$He  shielding in 
$^4$He, Ne and Ar  as 
gaseous solvents using exactly the same setup of the sample 
tube with the same set of capillaries. This series of measurements was 
specially designed for precise determination of the bulk susceptibility effects in our 
experiments, according to the approach discussed in section~\ref{sub5}.

\begin{figure}
\begin{center}
\includegraphics[scale=0.4]{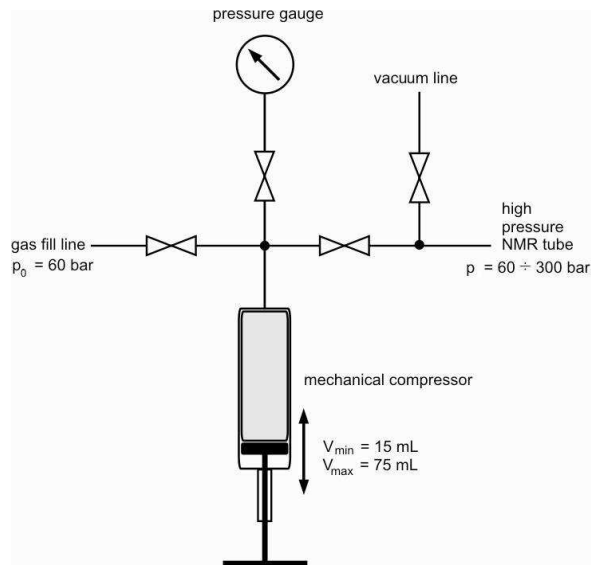}
\end{center}
\caption{A high pressure system for filling 
the zirconia NMR tube with a gas up to the pressure of 300 bar.}
\label{fig:spec}
\end{figure}

An efficient high-pressure system
built in our laboratory permitted
the NMR investigations of the hydrogen and helium gases for a wide range of densities. As indicated in Fig.~\ref{fig:spec}
the measurements in this system can be carried out continuously  up to
the total pressure of 300 bar.
All compartments were degassed when they were connected to
the vacuum line, then a small amount of the solute gas was supplied from
the same vacuum line and finally gas solvent was added and mechanically compressed. The pressure of gaseous solution was read by
the calibrated gauge and converted into the number of moles following
the van der Waals equation and appropriate coefficients for real
gases~\cite{handbook77}.
Gases: H$_2$ (Air Product, 99.9999\%), HD (Isotec, 98\% D), D$_2$ (Isotec, 99.96\%), $^3$He (Isotec, 99.96\%), $^4$He (Air Product, 99.9\%), Ne (Air Product, 99.999\%), Ar (Air Product, 99.9999\%)
and CO$_2$ (Aldrich, 99.8\%) from lecture bottles were used for
the preparation of samples without further purification.

We have performed all the
measurements of the $^1$H and $^2$H shielding for the hydrogen isotopomers, H$_2$, HD and D$_2$, as
a function of the solvent density where helium, neon and argon were used as
the solvents.
Comparing the $^1$H and $^2$H NMR signals from the H$_2$ and D$_2$ molecules we found that
the width at the half maximum of the deuterium signal is over an order of magnitude smaller than
the same parameter of protons in H$_2$, e.g.
$\Delta\nu_{1/2}$ = 86.9 Hz for H$_2$ in helium at 60 bar while
$\Delta\nu_{1/2}$ = 6.0 Hz for D$_2$ in helium at
the same pressure. In practice, this means that
the deuterium experiments deliver much more precise data for
the analysis than can be obtained from the $^1$H NMR observations of the H$_2$ molecule. Consequently,
we use  next
the $^2$H NMR experimental data for comparison of
the theoretical and experimental results.

For each discussed system, measurements have been performed for
more than 20 different solvent gas densities. In each case,
the linear fit represents
well the density dependence of the results,
with the
adjusted coefficient of determination larger than 0.995.
The experimental shielding constants were corrected for the gas imperfection, and
not only derived from the relations (\ref{eq:Bp}) and (\ref{rel}).

\section{Results and discussion}
\label{sec5}
\begin{figure}
\begin{tabular}{c}
(a) \includegraphics[scale=0.3,angle=270]{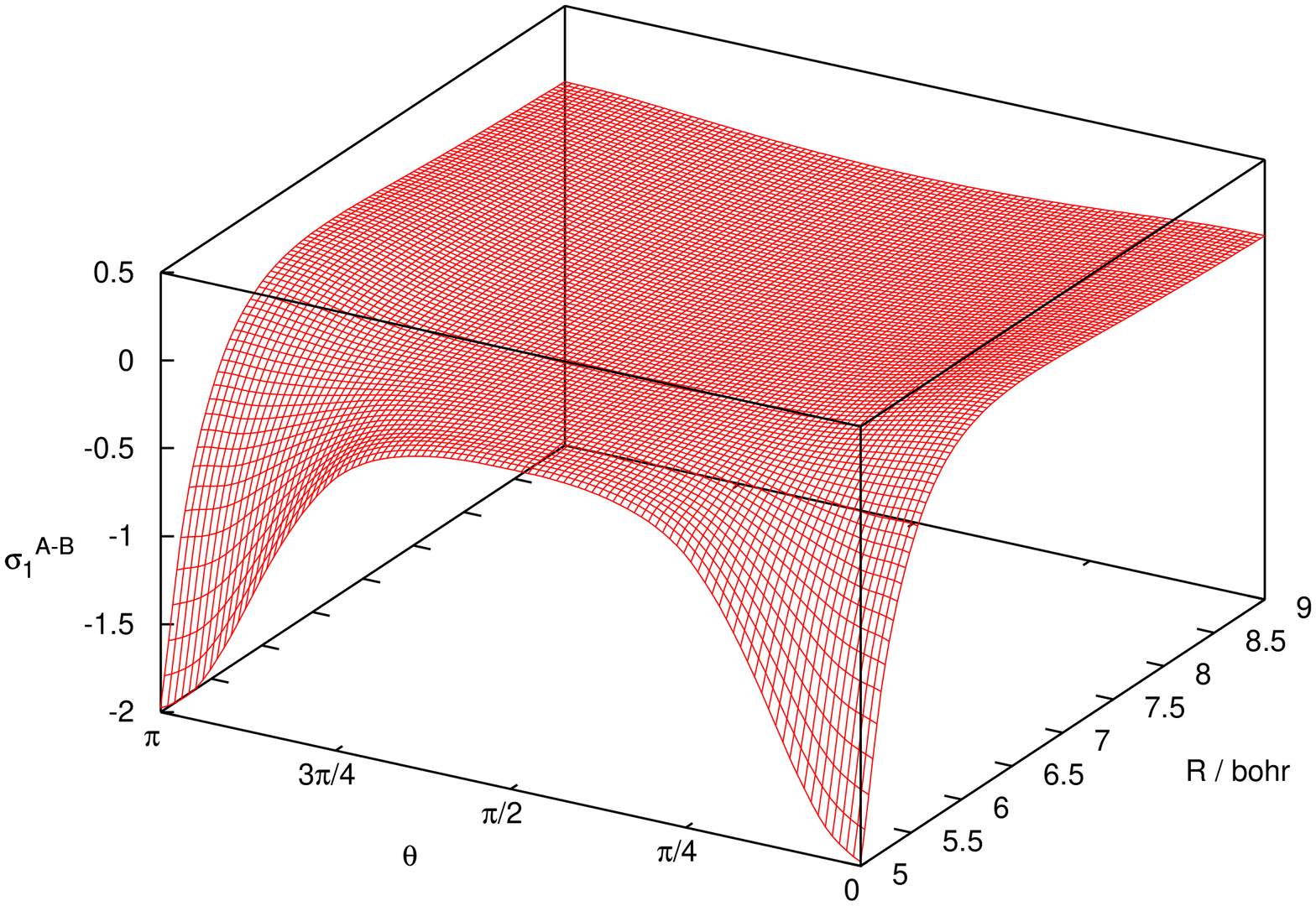} \\        
(b) \includegraphics[scale=0.3,angle=270]{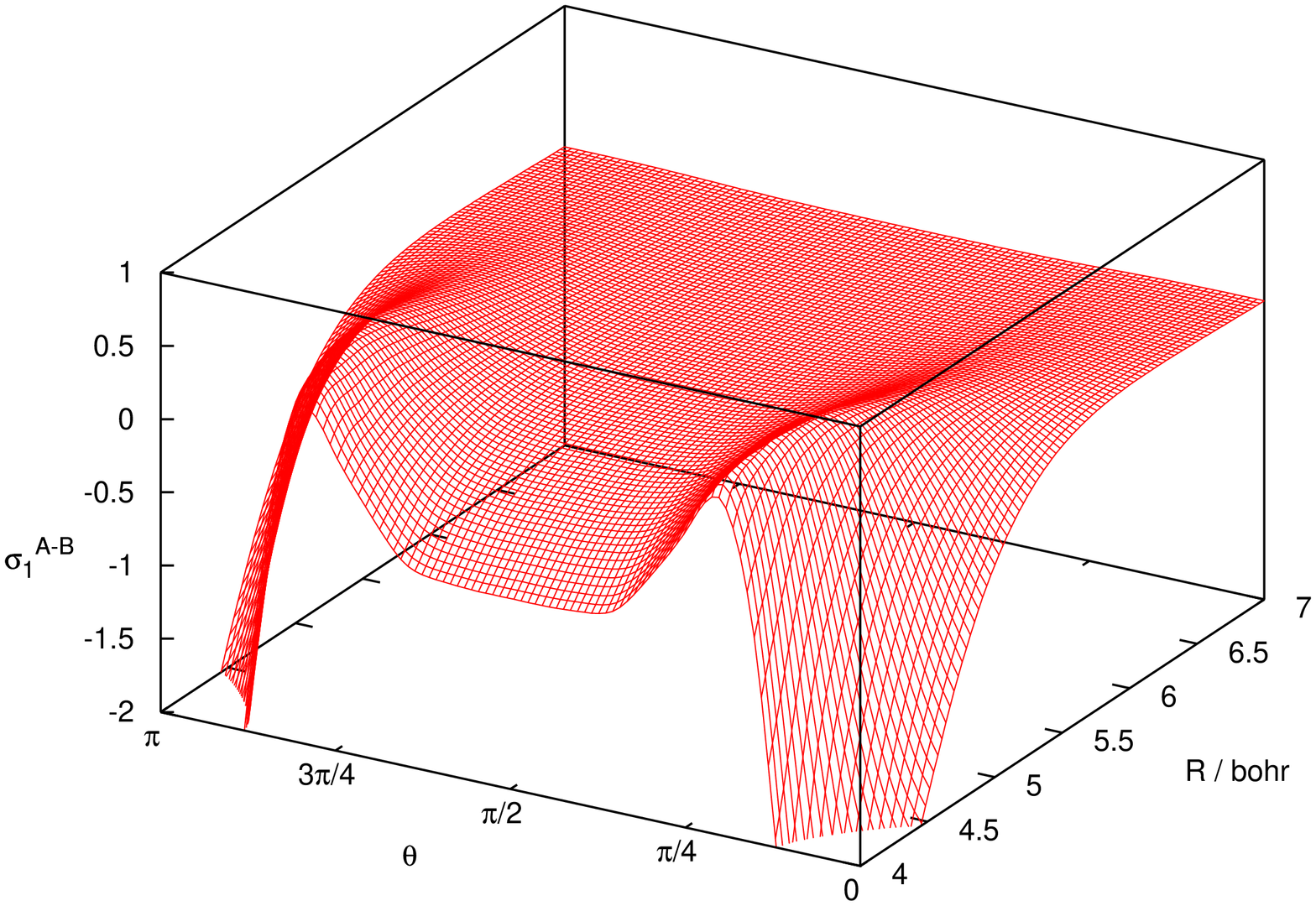} \\
{(c) \includegraphics[scale=0.3,angle=270]{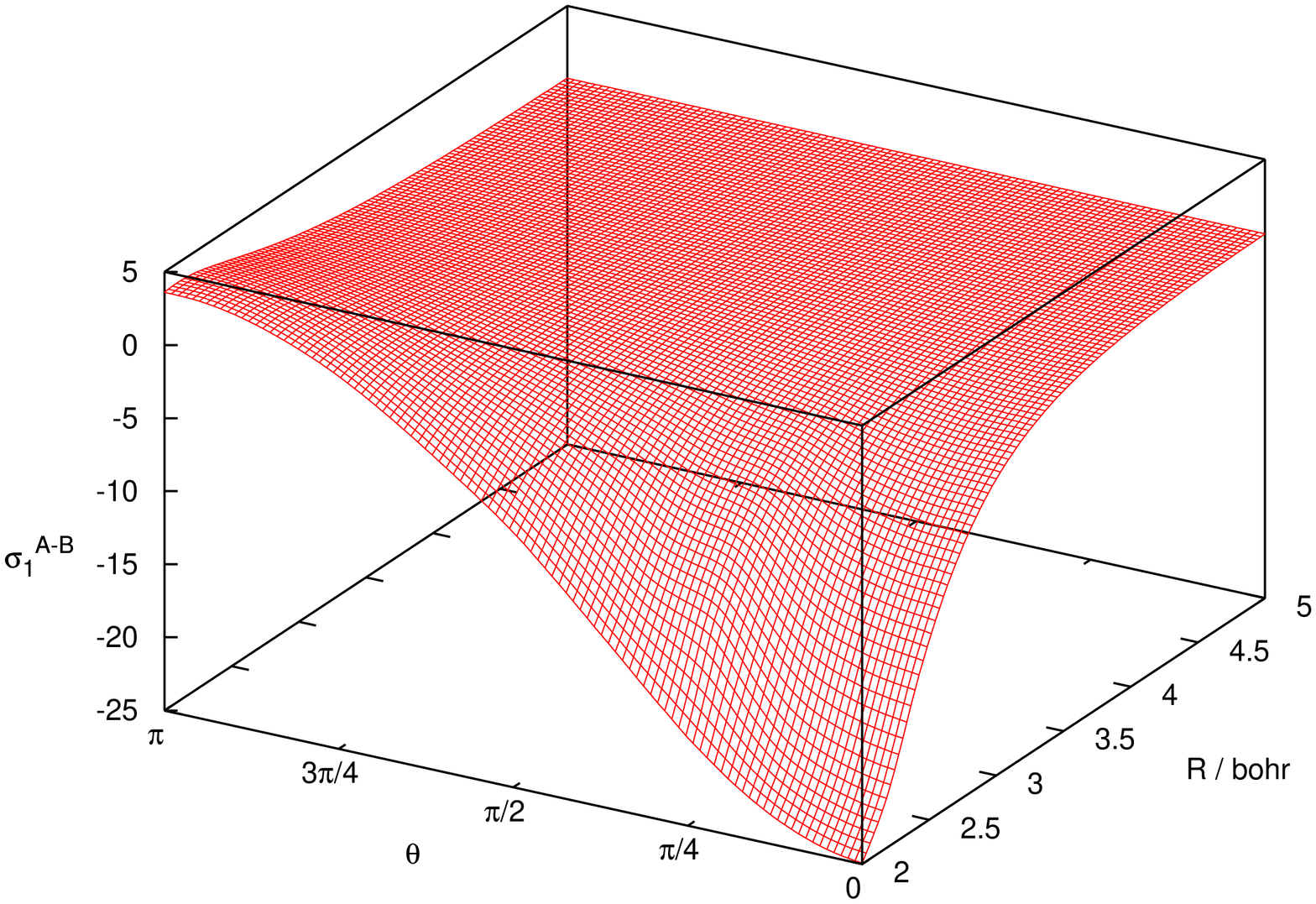}}
\end{tabular}
\caption{Geometry dependence of the interaction-induced shielding
constant for the 
(a) $^3$He--CO$_2$, (b) $^{13}$CO$_2$--He and  (c) H$_2$--He complexes.}
\label{shieldrt}
\end{figure}

We begin the discussion of the effects of intermolecular interactions on the shielding
constants with a brief summary of the {\em ab initio} results.
Three-dimensional plots of $\sigma_{\rm int}^{A-B}(X)$ for the H$_2$--He, 
$^3$He--CO$_2$, and $^{13}$CO$_2$--He complexes are  presented in Fig. \ref{shieldrt}. 
Similar plots for H$_2$--Ne
and H$_2$--Ar are not reported since their $R$ and $\theta$ dependence is nearly the
same as for H$_2$--He. An inspection of Fig. \ref{shieldrt} shows that 
$\sigma_{\rm int}^{H_2-He}$(H) does not show any strong variations on $R$ and $\theta$.
Only at very small intermolecular distances a stronger dependence shows up, but
at these geometries the interaction potential is strongly repulsive, so the
exponential Boltzmann factor is almost zero and these large variations  of 
$\sigma_{\rm int}^{H_2-He}$(H) do not contribute to $\sigma_{1}^{H_2-He}$(H).
Slightly more pronounced is the geometry dependence of the interaction-induced
shielding for both $^3$He--CO$_2$ and $^{13}$CO$_2$--He.
\begin{table}[ht]
\caption{Calculated {\em ab initio} values of $\sigma_1^{A-B}$ (ppm~mL/mol)}
\label{tab:temp}
\begin{center}
\begin{tabular}{c c c c }
\hline\hline
$T$ (K) & $\sigma_1^{H_2-He}$(H) & $\sigma_1^{H_2-Ne}$(H) & $\sigma_1^{H_2-Ar}$(H) \\
\hline
   150 & --0.324   & --0.230  &--4.071  \\
   200 & --0.352   & --0.281  &--4.025  \\
   250 & --0.381   & --0.330  &--4.085  \\
   280 & --0.398   & --0.358  &--4.144  \\
   300 & --0.410   & --0.377  &--4.189  \\
   320 & --0.422   & --0.396  &--4.237  \\
   350 & --0.440   & --0.424  &--4.314  \\
\hline
  & $\sigma_1^{He-CO_2}$(He)& $\sigma_1^{CO_2-He}$(C) \\
\hline
   150 & --6.579   &  1.3050    \\
   200 & --6.514   &  1.2894    \\
   250 & --6.548   &  1.2871    \\
   280 & --6.590   &  1.2876    \\
   300 & --6.622   &  1.2883    \\
   320 & --6.658   &  1.2889    \\
   350 & --6.714   &  1.2898    \\
\hline
\hline
\end{tabular}
\end{center}
\end{table}

Let us now analyse the temperature dependence of the 
$\sigma_1^{A-B}$ coefficients calculated from Eq.~(\ref{eq:Boltz}).
The results for all the systems are
shown in Table~\ref{tab:temp}.
An inspection of the Table
shows that
the temperature effects are
too small to be reliably determined from the experimental data 
(thus, in what follows
we shall only compare theoretical results with the experiment for $T$=300~K).
The dependence of the computed $\sigma_1^{A-B}$ on the temperature $T$ is almost
linear. It is interesting to note that for H$_2$--He, H$_2$--Ne, H$_2$--Ar, and
$^3$He--CO$_2$ systems $\sigma_1^{A-B}$ decreases with $T$, while for
$^{13}$CO$_2$--He the opposite is found. The values at the lowest temperature,
150~K, do not differ considerably from the room temperature data, suggesting 
that the quantum effects will start to play a noticeable role at still lower
temperatures.

The temperature dependence of
$\sigma_1^{A-B}$  was studied experimentally in glass samples 
as described earlier~\cite{kjmjbkmwjmr193}, but only for the $^3$He--CO$_2$ system
(rare gases like He, Ne and Ar could not in practice be used as solvents in such 
experiments).
Unfortunately, it was not possible to achieve sufficient precision
to perform a quantitative analysis of the results.    
In addition, a most important factor
required to determine 
$\sigma$($^3$He) in $^3$He--CO$_2$ as a function of the temperature---the
temperature dependence of CO$_2$ bulk susceptibility---is not known.

Before we consider the comparison of the {\em ab initio} and experimental results,
we recall that
for all the systems the bulk susceptibility corrections are
dominant.
The observed density dependence of $^2$H and $^3$He shielding in gaseous solutions
is shown in Fig.~\ref{figd2}
(the depicted range of densities
corresponds to the pressure of the solvent rare gas increasing up to 300 bar).
For comparison, we have
shown the effect of $\sigma_{{\rm 1bulk}}^B$ for Ar.
It is clear that for all the nuclei the total change of the shielding
in
the gas phase is very small, determined largely by $\sigma_{{\rm 1bulk}}^B$,
and
the precise evaluation of
the bulk susceptibility corrections
is crucial for the present study of weak molecular interactions.
The BSC effect can be neglected in
the analysis of the experimental data in two cases---when
a spherical NMR sample is prepared or
when the sample is fast spinning at
the magic angle. Unfortunately, neither of
these methods can
provide accurate results for compressed gas at high density.

Our final results, obtained for 300~K, are shown in Table~\ref{tab:sum}.
For each system, the BSC constitutes the essential part
of the measured effect, thus a minor error in the
evaluation of $\sigma_{{\rm 1bulk}}$ clearly leads to a very
significant error in $\sigma_1^{A-B}$. In particular,
the standard approximation, $\sigma_{{\rm 1bulk}} =  -({4\pi}/{3})\; \chi_{\rm M}$,
is not sufficiently accurate. 
The error bars shown in Table~\ref{tab:magn} and
Table~\ref{tab:sum} do not account for any systematic errors in the experiment, they
represent only the
errors of the linear fits to the observed density dependence of the results.
We have considered systematic errors 
arising from a limited 
precision of the nominal reading of the absolute frequency,
and errors in the control of the stability 
 of the external magnetic field; 
they 
limit the  precision of the measured shielding constants to $\pm$0.015 ppm.
However, let us recall that in many cases the determination of 
$\sigma_1^{A-B}(X)$ required two NMR experiments, one 
for the observed $A-B$ binary system and one for the $^3$He$-B$ solvent. 
Consequently, the error bars are at least doubled, to   
 $\pm$0.030 ppm for the $A-B$ system.
 Moreover, 
our experimental setup
 could slightly disturb the magnetic field as the sample was not 
 spinning during the measurements;
observing repeatedly the same samples 
we noticed deviations of 
 up to ±2 Hz 
in the measured frequencies. 
A complete analysis of these systematic errors, 
following 
the discussed precision of frequency measurements and possible 
 deviations in the gas density inside the NMR tube, gives 
 $\pm$0.50 ppm mL/mol
as an estimate of their contribution to the  error bars in 
 the $\sigma_1^{AB}$ values. This estimate does not 
include the tabulated errors of the linear fitting of the results, 
and does not take into account the left-over errors in the analysis of the
bulk susceptibility effects.

The errors in the {\em ab initio} calculations are also difficult to estimate.
The point-wise determined shielding surface is presumably accurate for the smallest
H$_2$--He system, the correlation and basis set errors becoming larger
for the other systems. Although the following stages---fitting the
potential and the property surfaces, followed by the Boltzmann average---appear to
be straightforward, approximations made in this part of the
calculation contribute significantly to the final error bars. As shown
in Fig.~\ref{shieldrt}, there are regions of the shielding surface of
opposite contributions to the induced shielding constant, and therefore
the final result depends heavily on a
significant cancellation of positive and negative contributions, which in turn
depends on the potential surface. Following various test calculations we
estimate that the errors of the computed  $\sigma_1^{A-B}$
should not exceed 15-20\% of the discussed above final {\em ab initio} values.

\begin{widetext}
\onecolumngrid
\begin{table}[ht]
\caption{{Measured and  calculated interaction--induced effects (ppm~mL/mol)}}
\label{tab:sum}
\begin{center}

\begin{tabular}{l d d d d}
\hline\hline
& \multicolumn{1}{c}{Measured} &  \multicolumn{1}{c}{Estimated} &  \multicolumn{1}{c}{Experimental} & \multicolumn{1}{c}{Calculated} \\
& \multicolumn{1}{c}{$\sigma^{AB}_{\rm 1}$} & \multicolumn{1}{c}{$\sigma_{{\rm 1bulk}}$}& \multicolumn{1}{c}{$\sigma_1^{A-B}$} & \multicolumn{1}{c}{$\sigma_1^{A-B}$} \\
\hline
\multicolumn{1}{l}{$\sigma$(D) in D$_2$--He} &
 8.07(8) &    8.62(12) &   \mbox{--}0.55(20) &\mbox{--}0.41  \\
\multicolumn{1}{l}{$\sigma$(D) in D$_2$--Ne} &
 29.83(3)&  29.56(6)  &  0.27(9) &\mbox{--}0.38  \\
\multicolumn{1}{l}{$\sigma$(D) in D$_2$--Ar} &
 76.44(32) & 80.26(30) & \mbox{--}3.78(62) & \mbox{--}4.19  \\
\multicolumn{1}{l}{$\sigma$(He) in $^3$He--CO$_2$} &
 84.7(24) & 93.22& \mbox{--}8.5(24) & \mbox{--}6.62  \\
\multicolumn{1}{l}{$\sigma$(C) in $^{13}$CO$_2$--He} &
  11.09(9) & 8.62(12)&  2.47(21) & 1.29  \\
\hline\hline
\end{tabular}
\end{center}
\end{table}
\end{widetext}

\section{Conclusions}
\label{sec6}
In this paper, we reported the first measurement of the changes of the NMR shielding
constants due to weak intermolecular interactions.
It became possible due to a new approach for the determination of bulk
susceptibility effects, which are dominant in the studied systems.
The
interpretation of the results is related to the corresponding
{\em ab initio} calculations, and we observe qualitative
agreement of the {\em ab initio} values with those derived from the
experimental data. There is a series of approximations that
should be analysed to improve this agreement.  In particular, it
is obvious that one cannot expect quantitative agreement without
a better description of the bulk susceptibility effects. We
have bypassed this problem transferring the necessary information
from one set of the experimental data---for pairs of rare gas atoms systems---
to another, that is to the studied molecule--atom systems.
Such an approach appears to yield satisfactory results in our case,
but in general a better theory, describing accurately the
bulk susceptibility corrections, is needed.
For larger systems, for instance involving
molecule-molecule interactions, the experiment may be easier,
but without a proper description of these effects the
interpretation of the results is almost impossible. Last but
not least, we note that the corresponding theoretical calculations
are also demanding, 
high level of the {\em ab initio} theory is required
to obtain a reliable description of the
small changes of the shielding constants due
to weak intermolecular forces.
For larger systems it may be difficult to achieve satisfactory
accuracy of the results, in particular when
the effects due to different parts of the shielding surface 
partially cancel out.

\section*{Acknowledgments}
We acknowledge support
of the
Polish Ministry of Science and Higher Education research grant
N~N204~244134 (2008-2011).
This project was partly
co-operated within the Foundation for Polish Science MPD
Programme co-financed by the EU European Regional Development Fund.

\bibliography{artikler,propert,mppubl,mjpubl,spinrev}
\bibliographystyle{unsrt}

\end{document}